# Fabrication of Hybrid Nanostructures via Nanoscale Laser-Induced Reshaping for Advanced Light Manipulation

Dmitry A. Zuev,[†,§] Sergey V. Makarov,[†,§] Valentin A. Milichko,[†] Sergey V. Starikov,[‡] Ivan S. Mukhin,[†,¶] Ivan A. Morozov,[¶] Ivan I. Shishkin,[†] Alexander E. Krasnok,[†,*] and Pavel A. Belov[†]

[†]Laboratory of Nanophotonics and Metamaterials, ITMO University, St. Petersburg 197101, Russia

[‡]Joint Institute for High Temperatures of the Russian Academy of Sciences, Moscow 125412, Russia

[¶]St. Petersburg Academic University, St. Petersburg 194021, Russia

§Contributed equally to this work

E-mail: krasnokfiz@mail.ru

The resonant metallic nanoparticles are proven to be efficient systems for the electromagnetic field control at nanoscale, owing to the ability to localize and enhance the optical field via excitation of strong plasmon resonances.[1–3] In turn, high index dielectric nanoparticles with low dissipative losses in the visible range, possessing magnetic and electric Mie-type resonances, offer a great opportunity for light control via designing of scattering properties.[4–8] Recently, the combination of these two paradigms in the form of metal-dielectric (hybrid) nanostructures (nanoantennas and metasurfaces) has allowed utilizing the advantages of both plasmonics and all-dielectric nanophotonics.[9] The hybrid nanostructures are prospective for beam steering,[10–12] optical switching,[13,14] high-harmonics generation,[15–17] directional emission,[18,19] engineering of local density of states,[19–21] magneto-optical activity,[22] ultrahigh optical absorption,[23] room-temperature laser emission,[24] and enhancement of photophysical effects.[9]

The unique properties of hybrid nanostructures mostly depend on the spectral overlapping of optical modes in the metal and dielectric nanoparticles. In particular, the precise positioning of plasmon resonance of the metal nanoparticle relative to magnetic and electric resonances of the dielectric one is crucial for control of a scattering power pattern and enhancement of electromagnetic field.[10,11,13,18,19]

The first feature of such hybrid nanostructures is the large difference between sizes of the metal and dielectric nanoparticles. The latter are usually several times larger than the former, because the Mie-resonances occur at the diameters larger than the wavelength in the dielectric nanoparticle.[25] The second feature is the necessity of application of different lithography methods and additional steps to create a metal-dielectric nanostructure with high precision. These requirements are strong limiting factors for fabrication of the asymmetrical (not a core-shell structure) hybrid nanostructures with certain optical properties. For example, the high-throughput wet-chemistry preparation can hardly be applied for fabrication of the hybrid nanostructures with precise control of optical properties, especially, based on high-index dielectric materials (Si, GaAs, etc.).[9,26] Laser ablation also gives random size distributions for the hybrid nanostructures.[11] Therefore, the multistage lithography[12,23,27] seems to be the most powerful tool to achieve desirable overlapping of the optical modes. However, due to the above mentioned specific requirements on size/position of individual components, the conventional lithography still suffers from the absence of flexibility.

The femtosecond (fs) laser melting at the nanoscale is proven to be an effective method for plasmonic nanoparticles reshaping.[28–33] As the direction of the melting-induced reshaping process is oriented from nanodisc to nanosphere due to irreversible action of surface tension, the plasmon resonance is to be shifted to the shorter wavelengths. However, the fs-laser melting has not been applied for hybrid nanostructures modification. In this work, we demonstrate for the first time a novel cost-effective approach for fabrication of ordered hybrid nanostructures consisting in femtosecond laser reshaping of metal-dielectric nanoparticles prepared via developed lithography methods. The lithographic stages provide special shape of the dielectric component to achieve the controllable modification, preserving both electric and magnetic optical Mie-type resonances, whereas local laser melting of individual hybrid nanoparticles allows selective reshaping of *only metal component* without affecting dielectric one. The presented method enables not only fabricating novel type of nonsymmetrical hybrid nanostructures, but *also tailoring their magnetic and electric optical resonances*. In particular, we demonstrate that the laser reshaping of a gold nanoparticle in an Au/Si nanodimer dramatically changes optical properties of the hybrid nanostructures. The proposed approach has a strong advantage over simple macroscopic heating owing to the possibility of high-precision modification of hybrid nanostructures.

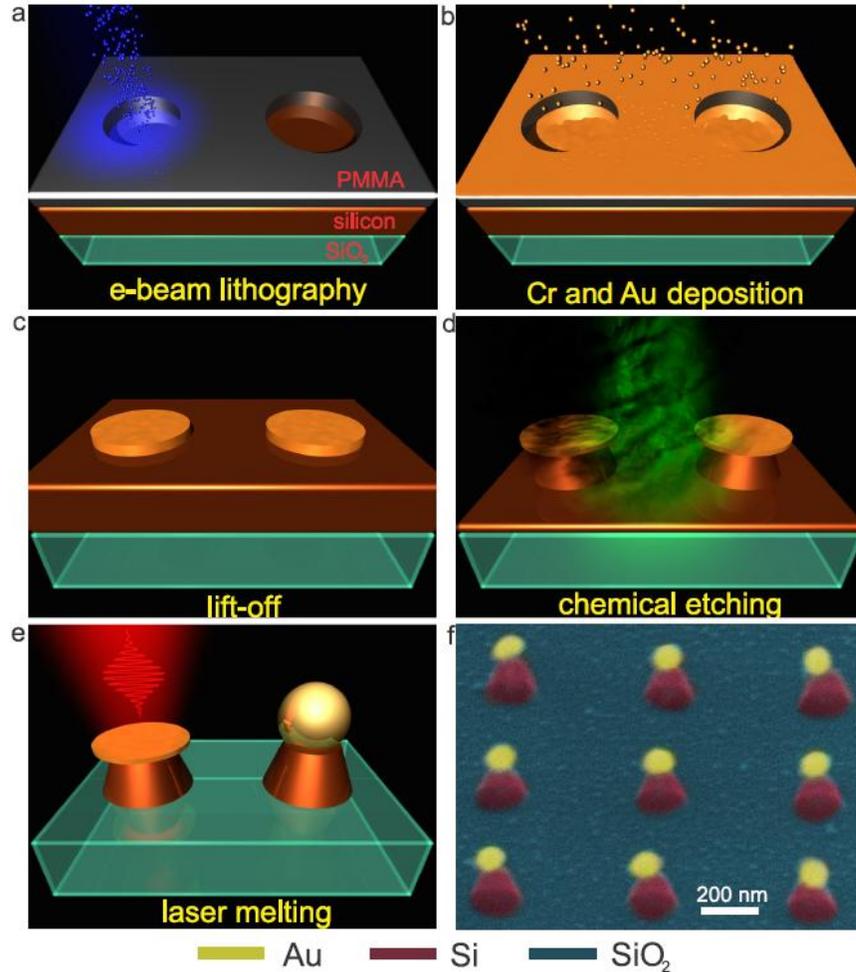

**Figure 1**: Hybrid nanostructures fabrication process. Lithography stages for Au nanodiscs fabrication via e-beam lithography using a positive resist (a), deposition of thin Cr (≈ 1 nm) and Au layers (b), and lift-off procedure (c). d) Gas phase chemical etching for Si nanocones fabrication. e) Laser post-processing stage: fs-laser reshaping of the Au nanodiscs for precise tuning of optical resonances in a hybrid nanodimer. f) SEM image represents the result of the Au nanodiscs reshaping to spherical nanoparticles. The Au nanoparticles are marked by yellow, the Si nanocones by crimson and the $SiO_2$ substrate by green colors.

We have fabricated the arrays of hybrid nanodimers and applied the femtosecond laser modification for optical properties tuning of both the whole array and individual nanodimers. First of all, the arrays of hybrid nanodimers (gold nanodisc placed on the top of truncated silicon nanocone) were produced by means of combination of e-beam lithography, metal evaporation, lift-off procedure, gas phase chemical etching and fs-laser irradiation as shown in **Figure 1**. At the first step, the *a*-Si:H layer with a thickness of ≈ 200 nm was deposited at a properly cleaned substrate of fused silica by the plasma enhanced chemical vapor deposition of $SiH_3$ gas. Then different arrays of metal nanodiscs consisted of the Cr/Au layers with thicknesses of ≈ 1 nm/10 nm, 1 nm/20 nm and 1 nm/30 nm were produced by means of e-beam lithography (**Figure 1**a), metal deposition (**Figure 1**b) followed by the lift-off procedures (**Figure 1**c). After that, the silicon layer was etched through fabricated metal mask using the radio frequency inductively coupled plasma technology in the presence of $SF_6$ and $O_2$ gases (**Figure 1**d). The etching was carried out with temperature of 265 K to fabricate Si nanostructures in the shape of nanocones. Finally, the fs-laser irradiation is applied for precise tuning of optical resonances in hybrid nanostructures via reshaping of the Au nanodiscs on the top of the Si nanocones (**Figure 1**e). The typical result of the hybrid nanodimers modification is represented in **Figure 1**f, demonstrating high repeatability of the fs-laser reshaping.

The series of experiments on the hybrid nanodimers modification were conducted on the identical arrays of the nanostructures distinguished only by the thickness ($h$) of the Au nanodisc (10, 20 and 30 nm). The arrays had the following geometrical parameters: the diameters of the Au nanodiscs ($d$) and base of Si nanocones ($h_c$) are ≈ 190±10 nm, the height of the Si nanocones is ≈ 200 nm, the period is about ≈ 600 nm, total size of each array is 17×17 μm. In the experiments laser fluence was varied in the range of 20–130 mJ/cm$^2$. The arrays of the structures were irradiated during scanning process. Therefore, the structures in the laser spot area were exposured by ∼ 10$^5$ pulses.

The general behavior of the structures modification process is similar for hybrid nanostructures with different thickness of the Au nanodisc. The final geometry of the hybrid nanodimer after modification is generally determined by the laser fluence and can be varied considerably. The results of the nanodimer shape modification are presented for the hybrid nanostructure with the Au nanodisc thickness of about 10 nm in **Figure 2**a. The slight modification of the Au nanodisc begins when the laser fluence ($F$) achieves the value of 28 mJ/cm$^2$. The following rise of the laser energy fluence up to 60 mJ/cm$^2$ leads to the Au nanodisc modification and then to the Au nanosphere creation on the top of the Si nanocone. The radius of the Au nanospheres ($R$) is directly proportional to $\sqrt[3]{h}$, because the volume of the Au nanoparticle remains constant during the laser-induced reshaping. The fluence exceeding the Au nanosphere modification threshold more than 10 mJ/cm$^2$ leads to melting of the Si nanocones on the boundary with Au nanospheres and shifting of the nanospheres from the centre to the side face of the Si nanocones. The laser pulses with fluences in the range $F \geq 100$ mJ/cm$^2$ strongly heat both the Au nanosphere and the Si nanocone, damaging the sphere-cone structure. In this fluence regime the spherical nanoparticles, presumably representing Si nanospheres covered by Au layer, emerge over the laser scan area as well as completely melted nanostructures.

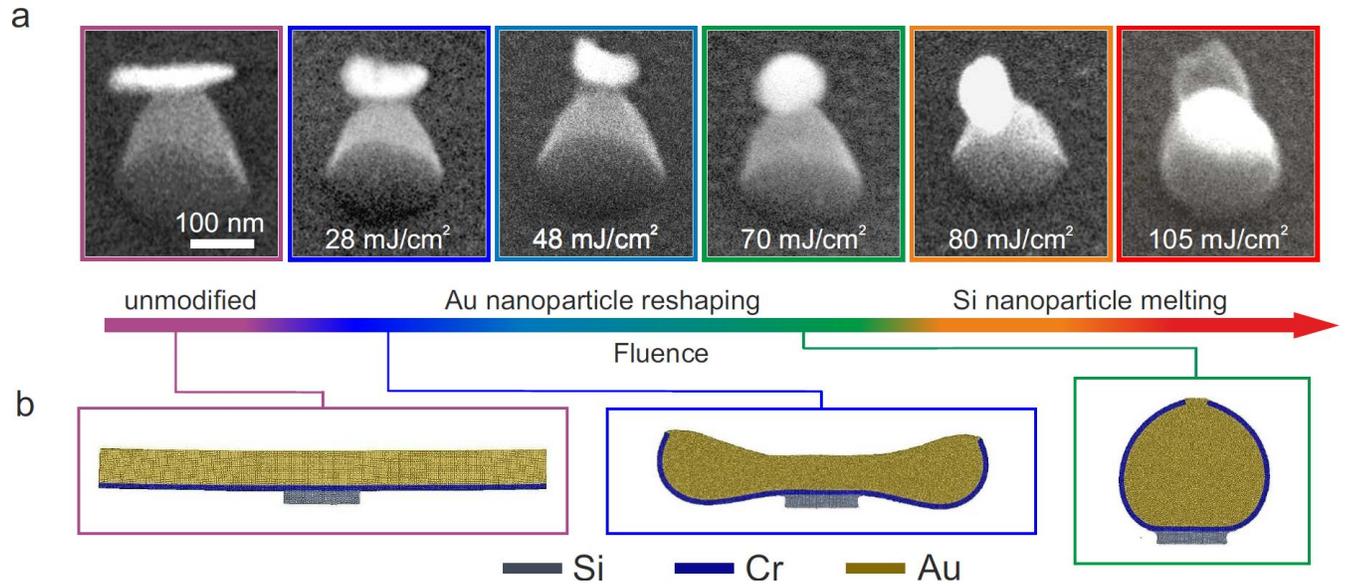

**Figure 2**: a) SEM image demonstrating modification of a hybrid nanodimer with Au nanodisc thickness *h* of 10 nm by fs-laser irradiation. The SEM images correspond to the typical structures in the following modification regimes: Au nanodisc deformation, transformation to Au nanospheres, Si nanocone melting, damage of the nanodimer. b) Molecular dynamics simulation of the Au nanoparticle reshaping. Grey color marks Si atoms; blue color - Cr atoms; yellow color - Au atoms.

The main mechanism, causing the change of the Au nanodisc shape at the intermediated fluences (lower than the Si nanocone melting threshold), can be described via dewetting process.[34,35] In this process, the surface morphology of a thin solid film changes with heating in order to minimize the surface energy. The dewetting is governed by temperature and width to thickness ratio of the heated nanoparticle.[36] In our experiments, the laser-induced reshaping depends on absorbed fluence and the diameter to thickness ratio (*d/h*) of the Au nanodiscs. To get insight into the nanodimers modification process and determine the *d* to *h* ratio for better engineering of hybrid nanodimer geometry and following precise reshaping, we have performed the molecular dynamics simulation, which is proven to be a powerful method to study the laser modification of matter.[37–41] The simulated system contains Au layer, Cr layer and several atomic layers of the Si support including the one fixed atomic layer for system fixation. The simulations take place in the quasi-two-dimensional case (the simulation cell size is about 4 nm and periodic boundary conditions are used). Such quasi-two-dimensional simulation may give a qualitative description of the matter evolution under laser irradiation (see discussion in Ref.[41]). The Au and Cr layers have identical length $d \approx 147$ nm and initial temperature T = 1350 K, the thicknesses of the layers are about 10 nm and 1 nm, respectively. **Figure 2**b shows the different possible configurations of the modelled Au nanodisc under continuous heating. In particular, the molecular dynamics simulation qualitatively reproduces the experimentally observed shapes of the modified Au nanodiscs at the intermediate stages. This evolution is determined by the changes of two energies: the surface energy of molten gold and the deformation/elastic energy of curved Cr layer. The simulation reveals that the thickness of the Cr layer ($h_{Cr}$) has an influence on anisotropy of the reshaped Au nanoparticle. While molten gold tends to minimize the surface area, the solid Cr layer with high temperature of melting (2130 K) prevents the system from the deformation. In this case, the parameter $h_{Cr}$ plays a crucial role, determining the deformation energy of the Cr layer. Our additional molecular dynamics simulation for different thicknesses of $h_{Cr}$ demonstrates, that the anisotropy of the formed nanostructures increases with the value of $h_{Cr}$ (see Supporting information, Section 1). Taking into account the results of the molecular dynamics simulation and the fact that the small thickness of $h_{Cr}$ results in reduction of the damping effect caused by the Cr layer on the resonance properties of plasmonic nanoparticles,[42] we reduced the thickness of $h_{Cr}$ as much as possible (down to 1 nm). The molecular dynamics simulation also provides the optimal value (≈ 14.7) of *d/h* ratio for the disc-to-sphere rershaping. In our experiments, the most appropriate Au layer thickness for the nanospheres formation is *h* ≈ 20 nm, corresponding to *d/h* ≈ 10. This value is in a good agreement with our simulations. The experimental data with different *d/h* values are demonstrated in Supporting information, Section 1.

The thickness of the Au nanodisc is also an important parameter in terms of the difference among energy fluences causing dramatical modification of the nanodimer shape. The studied laser fluence range can be divided into four fluence regimes (**Figure 3**a). The analysis of the regimes demonstrates, that one of the main advantages of such hybrid Au/Si nanostructures is the large difference between damage thresholds of the Au and Si components. Generally, gold has much lower melting temperature (1337 K) and enthalpy of fusion (12.55 kJ/mol) as compared with the corresponding values for silicon (1688 K, 50.21 kJ/mol), whereas their heat capacities are comparable. Therefore, such thermodynamics characteristics allow for selective reshaping of the Au component without affecting on the Si one. The first two regimes (I and II in **Figure 3**a) are related to the modification of the Au nanoparticle without influence on the Si nanocone. The areas III and IV in **Figure 3**a correspond to the Si nanocone modification and damage of the initial geometry of nanodimer, respectively. It was found that the hybrid nanostructures with the Au nanodisc thickness of 10 nm are more attractive for precise reshaping of the Au nanoparticle from the nanodisc to the nanosphere, possessing the biggest difference between the regimes I and II (≈50 mJ/cm$^2$), (**Figure 3**b). In turn, in terms of damage resistance, the most appropriate thickness of the Au nanodisc of the hybrid nanodimer corresponds to h ≈ 20 nm. These hybrid nanostructures have the biggest difference between the regimes II and III–IV, i.e. between 70 mJ/cm$^2$ and 100 mJ/cm$^2$, respectively (**Figure 3**a). It should be noted, that the fabricated Au nanospheres have a slight difference in *R*. Almost 90% of the radius values of nanoparticles are in the range of 90-100 nm (see Supporting information, Section 2).

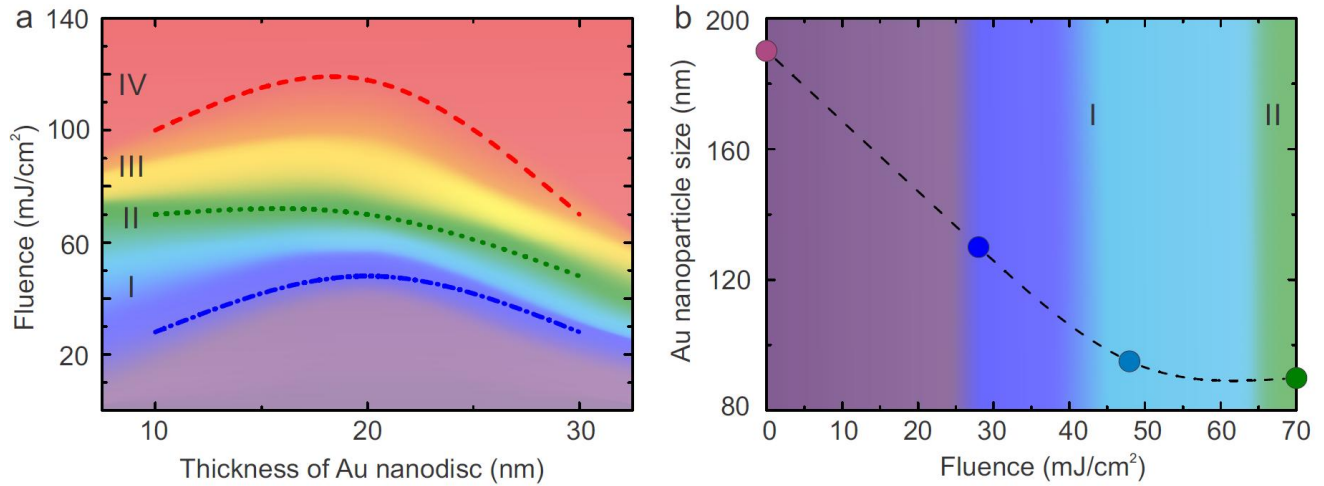

**Figure 3**: a) Thresholds of hybrid nanodimers reshaping with different thicknesses of the Au nanodiscs (10, 20 and 30 nm): Au nanodisc deformation - blue curve, transformation to Au nanospheres - green curve, Si nanocone melting - red curve. The following modification regimes are marked by: I - Au nanodisc deformation, II - transformation of Au nanodisc to Au nanospheres, III - Si nanocone melting, IV - damage of the nanodimer. b) Demonstration of precise control of the Au nanoparicle size in a hybrid nanodimer with *h* of 10 nm varying laser fluence.

The final optimization parameter is the shape of the Si nanoparticle. Since the top surface of the Si nanoparticle supports the Au nanoparticle during the laser-induced dewetting process, it strongly affects in intersurface interaction. In our experiments we reveal several advantages of a conical shape of the Si nanoparticle. Firstly, reduction of a supporting area for the Au nanodisc improves precision of the Au nanoparticle final position after dewetting. Secondly, increase of bottom surface of the Si nanoparticle enlarges a thermal contact area with a glass substrate, resulting in more effective cooling of the irradiated Si part and, consequently, better selectivity of the reshaping. Following these logic, we fabricated the Si nanoparticle possessing a truncated conical shape and minimized diameter of the top surface *a* ≤ 50 nm via etching at the relatively high substrate temperature (≈ 265 K). For larger diameters (*a* > 100 nm) we observe only shrinking of the Au nanodiscs under fs-laser heating down to the size, which is almost equal to the top surface diameter of the Si nanocone (see Supporting information, Section 3).

In order to demonstrate the precise manipulation of electric and magnetic responses of the hybrid nanostructures via fs-laser reshaping, we provide optical microscopy characterization of the hybrid nanodimers before and after the reshaping of the 20 nm thick Au nanodiscs, i.e. for the most optimal geometry, according to the above mentioned optimization.

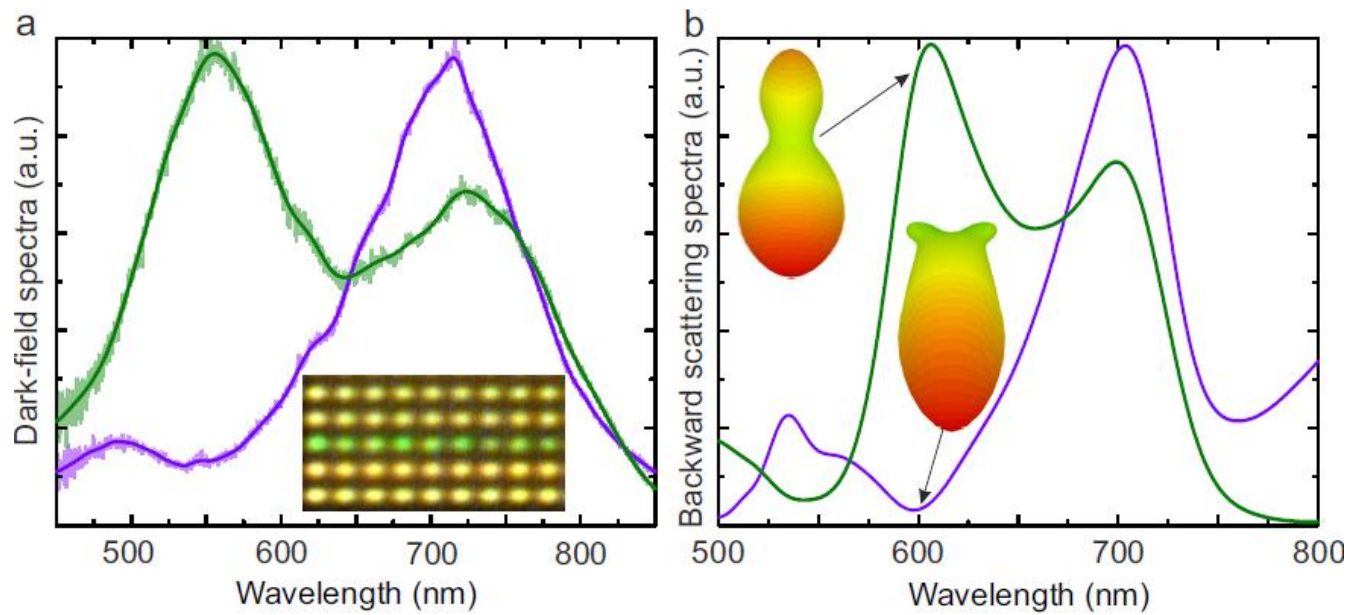

**Figure 4**: Tuning of scattering optical properties of resonance hybrid nanostructures. a) Scattering spectra measured from unmodified (violet curve) and modified (green curve) single hybrid nanodimers under p-polarized incident light. The inset shows the optical image of locally reshaped hybrid nanostructure in dark-field geometry for nonpolarized light. b) Numerically calculated scattering spectra/diagrams for an individual hybrid nanodimer representing Si nanocone ($a = 50$ nm, $h_c = 200$ nm) with Au nanodisc ($d = 190$ nm, $h = 20$ nm) or Au nanosphere ($R = 50$ nm).

The scattering spectra in dark-field scheme for the hybrid nanodimers before and after local fs-laser modification is demonstrated in **Figure 4**a. We observe the dramatic change of the spectra caused by the local fs-laser reshaping and resonant increase of scattering signal at the wavelength of about 550 nm (**Figure 4**a). The precise change of optical properties of hybrid nanostructure along individual rows is demonstrated by inset in **Figure 4**a. The tuning of scattering properties is in qualitative agreement with our numerical simulation (**Figure 4**b). The slight deviation of the experimental and numerical spectra can be explained by the absence of a substrate in the model, values of aperture of receiving and

Submitted to **ADVANCED MATERIALS**

focusing lenses, the differences in material parameters and dimensions, etc. The simulation of hybrid nanodimer scattering diagram shows the dramatic reconfiguration of the scattering pattern at the wavelengths of 600 nm (**Figure 4**b) after fs-laser reshaping. To explain this behavior, we have added the calculated scattering cross-section spectra for an individual Si nanoparticle (nanocone) and Au nanoparticle (nanodisc, nanocup and nanosphere) describing matching of spectral position of magnetic dipole and electrical dipole resonances of Si nanoparticle with electric dipole resonance of Au nanoparticle (see **Figure S5**a and **Figure S5**b in Supporting information, Section 4). The scattering spectrum of the Si nanocone with the specified size contains two peaks corresponding to electric (around 600 nm) and magnetic (around 700 nm) dipole resonances. Before the reshaping (i.e. in case of Au nanodisc) we have a resonant excitation of the electric dipole in the nanocone and nonresonant electric dipole excitation in the nanodisc at wavelength 600 nm. The absolute value of the electric dipoles are comparable, while they oscillate with a certain phase difference (not equal to $\pi$) that can be seen from electric field distribution at different times. The vector representation of electric field distribution is shown in **Figure S5**c (see Supporting information, Section 4). Such a system of two oscillating electric dipole moments forms an effective magnetic dipole moment of the entire hybrid nanodimer. However, since the electric dipoles oscillate with a phase difference, which is not equal to $\pi$, electrodynamic response of the entire system can be represented as the sum of the electric and magnetic dipole moments oscillating in phase. Such matching of the electric dipole and magnetic dipole oscillations increasing directivity of the nanodimer, is well known "Huygens source" regime.[43,44] Indeed, the unmodified hybrid nanodimer exhibits high directivity with almost zero backward scattering around the wavelength of 600 nm. After fs-laser reshaping of the Au nanodisc to the Au nanosphere the plasmon resonance shifts to the shorter wavelengths range and causes significant reconfiguration of the scattering pattern (see **Figure 4**b). Thus, the fs-laser reshaping of the hybrid nanostructures enables advanced (large and precise) manipulation of optical properties of hybrid nanostructures.

To conclude, we have demonstrated a novel approach for fabrication of hybrid nanostructures with magnetic optical response via femtosecond laser melting of asymmetrical metal-dielectric (Au/Si) nanoparticles created by lithographical methods. We have observed that the local laser melting enables selective reshaping of the metal components without affecting the dielectric ones. We have also revealed that the laser reshaping of the gold component in the Au/Si nanodimer allows to modify substantially the structure of the optical modes changing dramatically optical properties of the metal-dielectric nanostructures. The approach of the highly precise femtosecond laser melting at the nanoscale offers unique opportunities for fabrication of nanoantennas and metasurfaces consisting of asymmetrical hybrid nanostructures and manipulation of their optical properties. We believe that the represented results lay the groundwork of the fs-laser application for the large-scale fabrication of hybrid nanostructures and can be applied for effective light manipulation, biomedical and energy applications.

**Experimental Section**

*Fs-laser modification*: A commercial femtosecond laser system (Femtosecond Oscillator TiF–100F, Avesta Poject) was used, providing laser pulses at central wavelength of 790 nm (FWHM ≈ 11 nm), with a maximum pulse energy of 5 nJ, and a pulse duration of 100 fs at a repetition rate of 80 MHz. Laser energy was varied and controlled by an acousto-optical modulator (R23080–3–LTD, Gooch and Housego) and a power meter (FielfMax II, Coherent), respectively, while the pulse duration was measured by an autocorrelator (Avesta Poject). Laser pulses were focused by an objective (Olympus) with a numerical aperture (NA) of 0.75. Our measurements of dependence of the damage area size on laser energy carried out for the 100-nm Si thin film demonstrate, that the laser beam spot size is around 0.86 μm (See Supporting information, Section 5). The samples were placed on a three-dimensional air-bearing translating stage driven by brushless servomotors (ABL1000, Aerotech), allowing sample translation with accuracy around 0.1 μm.

*Samples characterization*: Preliminary optical imaging of the structures was provided immediately during the laser processing by integrated CCD camera. Scattering spectra measurements were carried out in a dark-field scheme, where the arrays irradiation was performed by p-polarized light from a halogen lamp (HL–2000–FHSA) at an angle of incidence of 70° with surface normal. Scattered signal collection was performed by means of objective Mitutoyo M Plan APO NIR (NA=0.7), which directed light to commercial spectrometer (Horiba LabRam HR) with CCD camera (Andor DU 420A–OE 325). Confocal optical scheme was optimized for signal collection from individual nanoparticles. A sketch of experimental setup for the polarization-resolved dark-field spectroscopy is represented in Supporting information, Section 6. The high-resolution morphology characterization was carried out by means of scanning electron microscope (SEM, Carl Zeiss, Neon 40).

*Numerical simulations of optical properties*: Properties of the hybrid nanostructures in the optical frequency range and scattering patterns have been studied numerically by using CST Microwave Studio. CST Microwave Studio is a full-wave 3D electromagnetic field solver based on finite-integral

Submitted to **ADVANCED MATERIALS**

time domain (FITD) solution technique. A nonuniform mesh was used to improve accuracy in the vicinity of the Au nanodiscs where the field concentration were significantly large. The dispersion model for Au and Si is taken from the works.[45–47]

*Numerical simulations of molecular dynamics*: The large scale molecular dynamics simulation was carried out using the LAMMPS code.[48] The AtomEye code was used for visualization of atomic dynamics.[49] For the description of gold we applied the EAM-potential[50] as a "cold part" of the electron-temperature-depended potential, which has been developed and successfully implemented in our previous works.[39,41] The interatomic interactions of Cr and Si atoms and cross-interactions are described by the Lennard–Jones potentials which are fitted to the melting temperatures (2130 K for Cr and 1688 K for Si). For the fitting procedure the rule $1.92 T_m = \varepsilon/k_B$ was used, where $\varepsilon$ and $k_B$ are the depth of the potential well and the Boltzmann constant, respectively. The $\sigma$-parameters of the Lennard–Jones potentials are equal to 0.26466 nm. The simulation box contains approximately $5 \cdot 10^5$ atoms. We heat the simulated system from room temperature up to 1350 K during 1 nanosecond using Langevin thermostat. This final temperature is 10% higher than the melting temperature of the used EAM-potential (about 1210 K). Full relaxation to the final state takes the time about 3–4 nanoseconds.

**Acknowledgements**

This work was supported by the Russian Science Foundation (Grant 15–19–00172). The part of the work related with the molecular dynamics simulation was partly supported by the President's Grant MK–7688.2015.8 (S.V.S.). The molecular dynamics calculations were carried out on the computer clusters MVS–100K of the Joint Supercomputer Center of RAS and "Lomonosov" of the Moscow State University. The authors are very thankful to Dr. Alexander S. Gudovskikh for PECVD deposition of amorphous silicon layer as well as Denis G. Baranov, Dr. Arseniy Kuznetsov and Professor Yuri S. Kivshar for helpful discussions and valuable advices.

**Competing financial interests**

The authors declare no competing financial interests.

**Contributions**

D.A.Z., S.V.M. and A.E.K. designed the experiments and analysed experimental results. I.S.M. and I.A.M. fabricated the hybrid nanostructures arrays and performed SEM characterization. D.A.Z. and

**Table of Contents**

Hybrid nanophotonics based on metal-dielectric nanostructures unifies the advantages of plasmonics and all-dielectric nanophotonics providing strong localization of light, magnetic optical response and specifically designed scattering properties. Here we demonstrate a novel approach for fabrication of ordered hybrid nanostructures via femtosecond laser melting of asymmetrical metal-dielectric (Au/Si) nanoparticles created by lithographical methods. The approach allows selective reshaping of the metal components of the hybrid nanoparticles without affecting dielectric ones. We apply the developed approach for tuning of the hybrid nanostructures scattering properties in the visible range. The experimental results are supported by molecular dynamics simulation and numerical solving of Maxwell's equations.

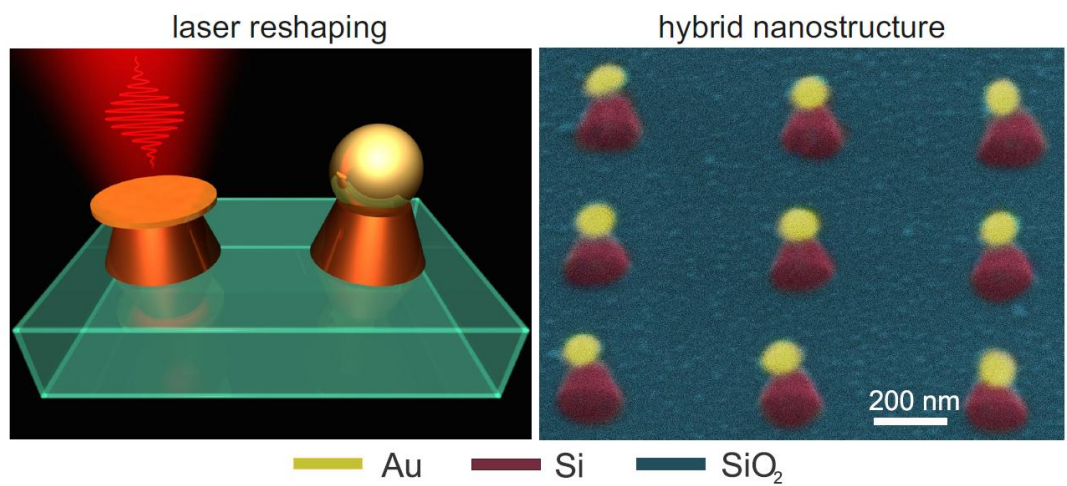